\documentclass[twocolumn,showpacs,pra,aps,superscriptaddress]{revtex4}

\usepackage{graphicx}
\usepackage{bm}
\usepackage{psfrag}
\usepackage{amsmath}
\usepackage{amssymb}
\usepackage{latexsym}
\usepackage{exscale}

\newcommand{\vect}[1]{\bm{#1}}
\newcommand{\ten}[1]{\mbox{\textbf{{\textsf{#1}}}}}

\newcommand{\sprod}{\!\cdot\!}
\newcommand{\tprod}{}
\newcommand{\vprod}{\!\times\!}
\newcommand{\trace}{\operatorname{Tr}}

\newcommand{\dif}{\mathrm{d}}
\newcommand{\mi}{\mathrm{i}} 
\newcommand{\me}{\mathrm{e}}

\begin{document}

\title{Universal scaling laws for dispersion interactions}

\author{Stefan Yoshi Buhmann}

\author{Stefan Scheel}
\affiliation{Quantum Optics and Laser Science, Blackett Laboratory,
Imperial College London, Prince Consort Road,
London SW7 2AZ, United Kingdom}

\author{James Babington}
\affiliation {Laboratoire de Physique et Mod\'{e}lisation des Milieux
Condens\'{e}s,\\ Universit\'{e} Joseph Fourier and CNRS, Maison des
Magist\`eres, BP 166, 38042 Grenoble, France.}

\date{\today}

\begin{abstract}
We study the scaling behaviour of dispersion potentials and forces
under very general conditions. We prove that a rescaling of an
arbitrary geometric arrangement by a factor $a$ changes the atom--atom
and atom--body potentials in the long-distance limit by factors
$1/a^7$ and $1/a^4$, respectively and the Casimir force per unit area
by $1/a^4$. In the short-distance regime, electric and magnetic bodies
lead to different scaling behaviours. As applications, we present
scaling functions for two atom--body potentials and display the
equipotential lines of a plate-assisted two-atom potential.
\end{abstract}

\maketitle

%%%%%%%%%%%%%%%%%%%%%%%%%%%%%%%%%%%%%%%%%%%%%%%%%%%%%%%%%%%%%%%%%%%%%%
%%%%%%%%%%%%%%%%%%%%%%%%%%%%%%%%%%%%%%%%%%%%%%%%%%%%%%%%%%%%%%%%%%%%%%

Scaling laws play a prominent role in the formulation of many physical
problems and occur naturally when studying critical phenomena in
particle physics, condensed matter and statistical mechanics. One
example is percolation theory~\cite{0876}, which has found
applications in understanding forest fires, oil-field extraction and
even measurement-based quantum computing \cite{QC}. 

Dispersion forces are effective quantum electromagnetic forces between
neutral, but polarisable objects \cite{0136,0853}. Casimir and Polder
found that dispersion forces are governed by simple power laws in the
long-distance limit \cite{0030}: The potential of a ground-state atom
at a distance $z_A$ from a perfectly conducting plate and that of two
ground-state atoms separated by a distance $r_{AB}$ are proportional
to $1/z_A^4$ and $1/r_{AB}^7$, respectively, while the force per unit
area between two perfectly conducting plates at separation $z$ follows
a $1/z^4$ law. Dispersion forces have since been studied for various
bodies of simple shapes such as semi-infinite half spaces \cite{0134},
plates of finite thickness \cite{0612,0019}, cylinders \cite{0066} and
spheres \cite{0077,0864}. In all of these cases, simple scaling laws
have been found for the long- and short-distance limits.

For electrostatic or gravitational interactions, power laws for the
forces between extended objects follow immediately by a volume
integration of $1/r$ potentials. Dispersion forces, on the contrary,
are due to an infinite hierarchy of microscopic $N$-point potentials
\cite{0087}, leading to a nontrivial geometry-dependence. Many-body
effects and the nontrivial dependence on geometry are at the heart of
current endeavours to gain a thorough theoretical \cite{0870} and
experimental understanding \cite{0555} of the Casimir effect and to
exploit it in nanotechnology applications \cite{0872}. The lack of
simple analytical solutions for dispersion forces in complex scenarios
necessitates general qualitative laws for what is achievable. Along
these lines, it has been proven that mirror-symmetric arrangements
always lead to attractive Casimir forces \cite{0503}, and duality
invariance has been established as a tool to study magnetoelectric
effects \cite{0827}. Scaling laws of the kind observed for simple
objects would be a powerful addition to this toolbox of general laws,
provided that they can be formulated beyond the special cases
mentioned above.

With this in mind, we will demonstrate in this Letter that for objects
of arbitrary shapes, dispersion interactions in the long- and
short-distance limits obey scaling laws; and we will identify the
respective scaling powers. Our proof relies on the known dependence of
dispersion potentials on the atomic polarisability $\alpha(\omega)$,
body permittivity $\varepsilon(\vect{r},\omega)$ and permeability
$\mu(\vect{r},\omega)$, where the latter determines the
electromagnetic Green tensor
\begin{equation}
\label{Sc3}
\biggl[\bm{\nabla}\vprod\frac{1}{\mu(\vect{r},\omega)}
 \bm{\nabla}\vprod
 \,-\,\frac{\omega^2}{c^2}\,\varepsilon(\vect{r},\omega)\biggr]
 \ten{G}(\vect{r},\vect{r}',\omega)
 =\bm{\delta}(\vect{r}-\vect{r}').
\end{equation}
In terms of these quantities, the Casimir--Polder (CP) potential of a
single electric ground-state atom and the van der Waals (vdW)
potential of two such atoms read
\begin{equation}
\label{Sc4}
U(\vect{r}_{A})=\frac{\hbar\mu_0}{2\pi}
 \int_0^\infty\dif\xi\,\xi^2\alpha_A(\mi\xi)\trace
 \ten{G}^{(1)}(\vect{r}_{A},\vect{r}_{A},\mi\xi),
\end{equation}
($\ten{G}=\ten{G}^{(0)}+\ten{G}^{(1)}$; bulk and scattering parts) and
\begin{multline}
\label{Sc5}
U(\vect{r}_{A},\vect{r}_{B})
 =-\frac{\hbar\mu_0^2}{2\pi}\int_0^\infty\dif\xi\,
 \xi^4\alpha_{A}(\mi\xi)\alpha_{B}(\mi\xi)\\[.5ex]
\times\trace[\ten{G}(\vect{r}_{A},\vect{r}_{B},\mi\xi)
 \sprod\ten{G}(\vect{r}_{B},\vect{r}_{A},\mi\xi)],
\end{multline}
respectively, and the Casimir force on a body of volume $V$ is given
by $\vect{F}=\int_{\partial V}\dif\vect{A}\sprod\ten{T}(\vect{r})$
with 
\begin{multline}
\label{Sc6}
\ten{T}(\vect{r})=-\frac{\hbar}{\pi}\int_{0}^{\infty} \dif\xi
 \biggl\{
 \biggl[\frac{\xi^2}{c^2}\,\ten{G}^{(1)}(\vect{r},\vect{r},\mi\xi)\\
+\bm{\nabla}\vprod\ten{G}^{(1)}(\vect{r},\vect{r},\mi\xi)
 \vprod\overleftarrow{\bm{\nabla}}'\biggr]
 -\frac{1}{2}\trace
 \biggl[\frac{\xi^2}{c^2}\,\ten{G}^{(1)}(\vect{r},\vect{r},\mi\xi)\\
 +\bm{\nabla}\vprod\ten{G}^{(1)}(\vect{r},\vect{r},\mi\xi)
 \vprod\overleftarrow{\bm{\nabla}}'\biggr]\ten{I}\biggr\}
\end{multline}
($\ten{I}$: unit tensor) \cite{0853}. We will first define the general
scaling problem and then solve it separately in the long- and
short-distance cases. 

%%%%%%%%%%%%%%%%%%%%%%%%%%%%%%%%%%%%%%%%%%%%%%%%%%%%%%%%%%%%%%%%%%%%%%
%%%%%%%%%%%%%%%%%%%%%%%%%%%%%%%%%%%%%%%%%%%%%%%%%%%%%%%%%%%%%%%%%%%%%%

\paragraph{The scaling problem.}

\begin{figure}[!t!]
\noindent\vspace*{-2ex}
\begin{center}
\includegraphics[width=\linewidth]{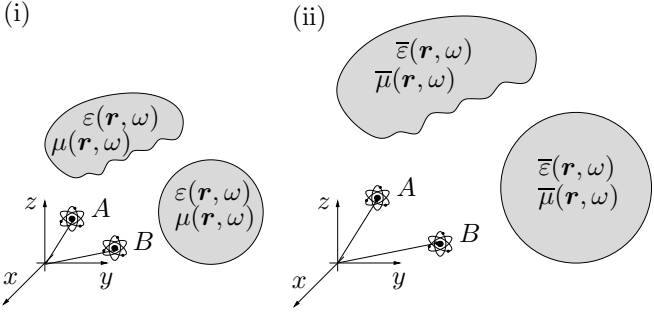}
\end{center}\vspace*{-2ex}
\caption{
\label{Fig1}
(i) Original and (ii) scaled configurations of bodies and atoms
($a=1.4$).
}\vspace*{-2ex}
\end{figure}

Consider an arbitrary arrangement of linearly responding bodies
characterised by their permittivity $\varepsilon(\vect{r},\omega)$ and
permeability $\mu(\vect{r},\omega)$, with one or two atoms at
positions $\vect{r}_A$ and $\vect{r}_B$ [Fig.~\ref{Fig1}(i)]. The
scaled arrangement (scaling factor $a\neq 0$) is described by the
permittivity and permeability
\begin{equation}
\label{Sc7}
\overline{\varepsilon}(\vect{r},\omega)
 =\varepsilon(\vect{r}/a,\omega),\qquad
\overline{\mu}(\vect{r},\omega)
 =\mu(\vect{r}/a,\omega);
\end{equation}
with the atomic positions being scaled accordingly:
$\overline{\vect{r}}_A=a\vect{r}_A$,
$\overline{\vect{r}}_B=a\vect{r}_B$ [Fig.~\ref{Fig1}(ii)].
%%%%%%%%%%%%%%%  F I G U R E %%%%%%%%%%%%%%%%%%%%%%%%%%%%%%%%%%%%%%%%%
%%%%%%%%%%%%%%%%%%%%%%%%%%%%%%%%%%%%%%%%%%%%%%%%%%%%%%%%%%%%%%%%%%%%%%
%
%%%%%%%%%%%%%%%%%%%%%%%%%%%%%%%%%%%%%%%%%%%%%%%%%%%%%%%%%%%%%%%%%%%%%%
%%%%%%%%%%%%%%%%%%%%%%%%%%%%%%%%%%%%%%%%%%%%%%%%%%%%%%%%%%%%%%%%%%%%%%
%
\paragraph{Interactions for long distances.}
We speak of the long-distance regime when all distances are much
larger than the wavelengths of the atomic and medium response
functions. In this case, we can approximate the latter by their static
values, $\alpha(\omega)\simeq\alpha$,
$\varepsilon(\vect{r},\omega)\simeq\varepsilon(\vect{r})$,  
$\mu(\vect{r},\omega)\simeq\mu(\vect{r})$, so the Green tensor is
determined by
\begin{equation}
\label{Sc8}
\biggl[\bm{\nabla}\vprod\frac{1}{\mu(\vect{r})}
 \bm{\nabla}\vprod
 \,-\,\frac{\omega^2}{c^2}\,\varepsilon(\vect{r})\biggr]
 \ten{G}(\vect{r},\vect{r}',\omega)
 =\bm{\delta}(\vect{r}-\vect{r}').
\end{equation}
The Green tensor of the scaled arrangement obeys
\begin{equation}
\label{Sc9}
\biggl[\bm{\nabla}\vprod\frac{1}{\overline{\mu}(\vect{r})}
 \bm{\nabla}\vprod
 \,-\,\frac{\omega^2}{c^2}\,\overline{\varepsilon}(\vect{r})\biggr]
 \overline{\ten{G}}(\vect{r},\vect{r}',\omega)
 =\bm{\delta}(\vect{r}-\vect{r}').
\end{equation}
By renaming $\vect{r}\mapsto a\vect{r}$, $\omega\mapsto\omega/a$ and
using Eq.~(\ref{Sc7}) and $\delta(a\vect{r})=\delta(\vect{r})/a^3$,
we find that
\begin{equation}
\label{Sc10}
\biggl[\bm{\nabla}\vprod\frac{1}{\mu(\vect{r})}
 \bm{\nabla}\vprod
 \,-\,\frac{\omega^2}{c^2}\,\varepsilon(\vect{r})\biggr]
 a\overline{\ten{G}}(a\vect{r},a\vect{r}',\omega/a)
 =\bm{\delta}(\vect{r}-\vect{r}').
\end{equation}
Comparison with Eq.~(\ref{Sc8}) reveals the scaling
\begin{equation}
\label{Sc11}
\overline{\ten{G}}(a\vect{r},a\vect{r}',\omega/a)
=(1/a)\ten{G}(\vect{r},\vect{r}',\omega).
\end{equation}
Substitution into the CP potential~(\ref{Sc4}) leads to
\begin{multline}
\label{Sc12}
\overline{U}(a\vect{r}_{A})=\frac{\hbar\mu_0\alpha_A}{2\pi}
 \int_0^\infty\dif\xi\,\xi^2\,\trace
 \overline{\ten{G}}^{(1)}(a\vect{r}_{A},a\vect{r}_{A},\mi\xi)\\
=\frac{\hbar\mu_0\alpha_A}{2\pi}
 \int_0^\infty\frac{\dif\xi}{a}\,\frac{\xi^2}{a^2}\,\,
\trace
 \overline{\ten{G}}^{(1)}(a\vect{r}_{A},a\vect{r}_{A},\mi\xi/a)\\
=(1/a^4)U(\vect{r}_{A}).
\end{multline}
The CP force thus scales as
$\overline{\vect{F}}(a\vect{r}_{A})=(1/a^5)\vect{F}(\vect{r}_{A})$. 
Analogously, Eq.~(\ref{Sc11}) can be used to derive the 
following scaling laws for the vdW potential~(\ref{Sc5}) and the Casimir force
per unit area~(\ref{Sc6}):
\begin{gather}
\label{Sc13}
\overline{U}(a\vect{r}_A,a\vect{r}_B)
=(1/a^7)U(\vect{r}_A,\vect{r}_B),\\
\label{Sc14}
\overline{\ten{T}}(a\vect{r})
=(1/a^4)\ten{T}(\vect{r}),
\end{gather}
so the total Casimir force behaves as $\overline{\vect{F}}
=(1/a^2)\vect{F}$.
%
%%%%%%%%%%%%%%%%%%%%%%%%%%%%%%%%%%%%%%%%%%%%%%%%%%%%%%%%%%%%%%%%%%%%%%
%%%%%%%%%%%%%%%%%%%%%%%%%%%%%%%%%%%%%%%%%%%%%%%%%%%%%%%%%%%%%%%%%%%%%%
%
\paragraph{Interactions for short distances.}
In the short-distance or nonretarded re\-gime, all distances are much
smaller than the characteristic atomic and medium wavelengths. A
simple example shows that here no universal scaling law exists in the
general case: The nonretarded CP potential of an atom at distance
$z_A$ from a magnetoelectric half space reads \cite{0019}
\begin{equation}
\label{Sc15}
U(z_{A})=-\frac{C_3}{z_A^3}+\frac{C_1}{z_A},
\end{equation}
which is incompatible with a relation of the form~(\ref{Sc12}).

However, scaling laws can still be formulated by distinguishing
between purely electric and purely magnetic environments. For purely
electric bodies, the Green tensor~(\ref{Sc3}) can be given by the
Dyson equation \cite{0853}
\begin{multline}
\label{Sc16}
\ten{G}(\vect{r},\vect{r}',\omega)
=\ten{G}^{(0)}(\vect{r},\vect{r}',\omega)\\
 +\frac{\omega^2}{c^2}
 \int\dif^3s\,\chi(\vect{s},\omega)
 \ten{G}^{(0)}(\vect{r},\vect{s},\omega)
 \sprod\ten{G}(\vect{s},\vect{r}',\omega)
\end{multline}
($\chi=\varepsilon-1$) where
\begin{multline}
\label{Sc17}
\ten{G}^{(0)}(\vect{r},\vect{r}',\omega)
 =-\frac{c^2\me^{\mi\omega\rho/c}}{4\pi\omega^2\rho^3}
 \biggl\{\biggl[1-\mi\,\frac{\omega\rho}{c}
 -\Bigl(\frac{\omega\rho}{c}\Bigr)^2\biggr]\ten{I}\\
-\biggl[3-3\mi\,\frac{\omega\rho}{c}
 -\Bigl(\frac{\omega\rho}{c}\Bigr)^2\biggr]
 \vect{e}_\rho\tprod\vect{e}_\rho\biggr\}
-\frac{c^2}{3\omega^2}\,\bm{\delta}(\vect{\rho})
\end{multline}
($\bm{\rho}=\vect{r}-\vect{r}'$) is the free-space Green tensor. In
the short-distance limit $\omega\rho/c\ll 1$, the latter reduces to
\begin{equation}
\label{Sc17b}
\ten{G}^{(0)}(\vect{r},\vect{r}',\omega)
=-\frac{c^2[\ten{I}-3\vect{e}_\rho\tprod\vect{e}_\rho]}
 {4\pi\omega^2\rho^3}
-\frac{c^2}{3\omega^2}\,\bm{\delta}(\bm{\rho}).
\end{equation}
Starting from the analogous Dyson equation for the scaled Green
tensor, we make the substitutions $\vect{r}\mapsto a\vect{r}$,
$\vect{r}'\mapsto a\vect{r}'$ and $\vect{s}\mapsto a\vect{s}$. After
using Eq.~(\ref{Sc7}) and the scaling
$\ten{G}^{(0)}(a\vect{r},a\vect{r}',\omega)%
=(1/a^3)\ten{G}^{(0)}(\vect{r},\vect{r}',\omega)$ of
Eq.~(\ref{Sc17b}), a comparison with (\ref{Sc16}) reveals the scaling
\begin{equation}
\label{Sc18}
\overline{\ten{G}}(a\vect{r},a\vect{r}',\omega)
=(1/a^3)\ten{G}(\vect{r},\vect{r}',\omega)
\end{equation}
for the full Green tensor. Substitution into
Eqs.~(\ref{Sc4})--(\ref{Sc6}) immediately implies the scaling laws
\begin{gather}
\label{Sc19}
\overline{U}(a\vect{r}_A)
=(1/a^3)U(\vect{r}_A),\\
\label{Sc20}
\overline{U}(a\vect{r}_A,a\vect{r}_B)
=(1/a^6)U(\vect{r}_A,\vect{r}_B),\\
\label{Sc21}
\overline{\ten{T}}(a\vect{r})
=(1/a^3)\ten{T}(\vect{r})
\end{gather}
where we have used the fact that $\ten{G}^{(1)}$ dominates over
$\bm{\nabla}\vprod\ten{G}^{(1)}\vprod\overleftarrow{\bm{\nabla}}'$ in
the short-distance limit. 

For an arrangement of purely magnetic bodies, the nonretarded Green
tensor obeys the Dyson equation
\begin{multline}
\label{Sc22}
\ten{G}^{(1)}(\vect{r},\vect{r}',\omega)
 =\int\dif^3s\,\zeta(\vect{s},\omega)
 \bigl[\ten{G}^{(0)}(\vect{r},\vect{s},\omega)\vprod
 \overleftarrow{\vect{\nabla}}_{\!\vect{s}}\bigr]\\
 \sprod\bigl[\vect{\nabla}_{\!\vect{s}}\vprod
 \ten{G}^{(0)}(\vect{s},\vect{r}',\omega)
+\vect{\nabla}_{\!\vect{s}}\vprod
 \ten{G}^{(1)}(\vect{s},\vect{r}',\omega)\bigr]
\end{multline}
($\zeta=1/\mu-1$) with nonretarded free-space Green tensors
\begin{equation}
\label{Sc24}
\vect{\nabla}\vprod\ten{G}^{(0)}(\vect{r},\vect{r}',\omega)
 =-\ten{G}^{(0)}(\vect{r},\vect{r}',\omega)
 \vprod\overleftarrow{\vect{\nabla}}'
 =-\frac{\vect{e}_\rho\vprod\ten{I}}
 {4\pi\rho^2}\,.
\end{equation}
We read off $(1/a^2)$ scalings for $\vect{\nabla}\vprod\ten{G}^{(0)}$
and $\ten{G}^{(0)}\vprod\overleftarrow{\vect{\nabla}}'$, so
following similar steps as above, the Dyson equation~(\ref{Sc22})
together with Eq.~(\ref{Sc7}) implies
\begin{equation}
\label{Sc25}
\overline{\ten{G}}{}^{(1)}(a\vect{r},a\vect{r}',\omega)
=(1/a)\ten{G}^{(1)}(\vect{r},\vect{r}',\omega).
\end{equation}
Using Eq.~(\ref{Sc4}), the nonretarded CP potential scales as
\begin{equation}
\label{Sc26}
\overline{U}(a\vect{r}_A)
=(1/a)U(\vect{r}_A)
\end{equation}
for purely magnetic bodies. The vdW potential~(\ref{Sc5}) contains
contributions from the bulk and scattering Green tensors with
different scalings. We separate it into a free-space part $U^{(0)}$
that contains only $\ten{G}^{(0)}$ and scales according to
Eq.~(\ref{Sc20}) and a body-induced part $U^{(1)}$. The latter is
dominated by the mixed terms $\ten{G}^{(0)}\ten{G}^{(1)}$ for purely
magnetic bodies in the short-distance limit; it scales as
\begin{equation}
\label{Sc27}
\overline{U}{}^{(1)}(a\vect{r}_A,a\vect{r}_B)
=(1/a^4)U^{(1)}(\vect{r}_A,\vect{r}_B).
\end{equation}
The Casimir force~(\ref{Sc6}) is dominated by
$\vect{\nabla}\vprod\ten{G}^{(1)}%
\vprod\overleftarrow{\vect{\nabla}}'$ with its $(1/a^3)$ scaling for
purely magnetic bodies, so that
\begin{equation}
\label{Sc28}
\overline{\ten{T}}(a\vect{r})
=(1/a^3)\ten{T}(\vect{r}).
\end{equation}
%
%%%%%%%%%%%%%%%%%%%%%%%%%%%%%%%%%%%%%%%%%%%%%%%%%%%%%%%%%%%%%%%%%%%%%%
%%%%%%%%%%%%%%%%%%%%%%%%%%%%%%%%%%%%%%%%%%%%%%%%%%%%%%%%%%%%%%%%%%%%%%
%
%%%%%%%%%%%%%%%  F I G U R E %%%%%%%%%%%%%%%%%%%%%%%%%%%%%%%%%%%%%%%%%
\begin{figure}[!t!]
\noindent\vspace*{-2ex}
\begin{center}
\includegraphics[width=0.7\linewidth]{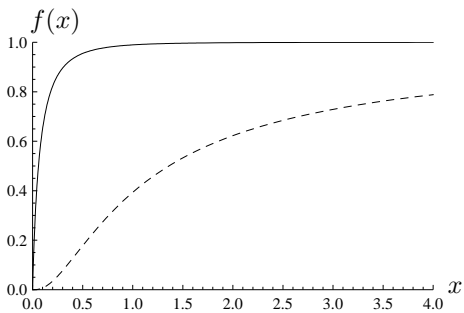}
\end{center}\vspace*{-2ex}
\caption{
\label{Fig2}
Scale functions for the potential of an atom in front of a Si plate
(solid line) and near a perfectly conducting sphere (dashed line).
}\vspace*{-2ex}
\end{figure}%
%%%%%%%%%%%%%%%%%%%%%%%%%%%%%%%%%%%%%%%%%%%%%%%%%%%%%%%%%%%%%%%%%%%%%%
\paragraph{Applications.}
In the simplest situations where dispersion forces depend on a single
distance parameter, the scaling laws directly determine the dependence
on that parameter. For instance, the long-distance scaling laws
(\ref{Sc12})--(\ref{Sc14}) imply the power laws for dispersion
interactions involving atoms and perfectly conducting plates
mentioned above. 

For a class of geometries involving a distance parameter $z$ and a
single size parameter $d$, the scaling laws can be employed to write
potentials and forces in the form $(C_k/z^k)f(d/z)$, all relevant
information being contained in the scaling function $f(x)$. In
Fig.~\ref{Fig2} we display the scaling functions for the potential of
an atom at distance $z_A$ from a Si plate of thickness $d$
($x=d/z_A$) in the long-distance limit \cite{0019} and for the
nonretarded potential of a perfectly conducting sphere of radius $R$
($x=R/z_A$) \cite{0864}.

The plate potential reaches its half-plate limit with associated 
$1/z_A^4$ asymptote already for $x=d/z_A\gtrsim 0.5$, showing that
finite thickness effects can be neglected for moderately thick plates
even for dielectrics. For very thin plates with $x=d/z_A\lesssim
0.1$, the scale function of the plate is linear for small $x$,
implying a $x/z_A^4\propto 1/z_A^5$ potential. A rather abrupt change
between the two power laws occurs between the two extremes.

The scale function of the sphere saturates much more slowly to its
large-$x$ asymptote where a $1/z_A^3$ half-space potential is
observed. This indicates that proximity force approximations
\cite{0601} should be used with care. The scale function of the sphere
potential is cubic for small $x$, corresponding to a $x^3/z_A^3\propto
1/z_A^6$ asymptote.

\begin{figure}[!t!]
\noindent\vspace*{-2ex}
\begin{center}\vspace*{-2ex}
\includegraphics[width=\linewidth]{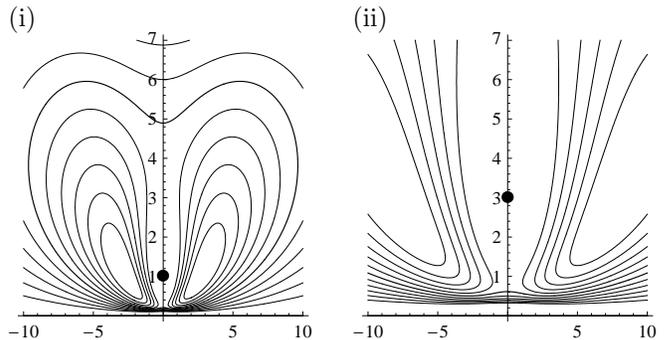}
\end{center}
\caption{
\label{Fig3}
Retarded vdW potential next to a perfectly conducting plate. Atom $B$
is held at different fixed positions (large dot). The thick contour
denotes $U/U^{(0)}=1$, values are increasing towards the exterior of
this contour in steps of $0.02$.
}\vspace*{-2ex}
\end{figure}

As a more complex example, we consider the vdW potential of two atoms
$A$ and $B$ in front of a perfectly conducting plate in
the long-distance limit \cite{0679}. In Fig.~\ref{Fig3}(i), we show
the plate-induced enhancement of the potential with respect to its
free space value for a given distance of atom $B$ from the plate. The
results for a different distance can then be obtained from a scaling
transformation, cf.~Fig.~\ref{Fig3}(ii). 
%%%%%%%%%%%%%%%  F I G U R E %%%%%%%%%%%%%%%%%%%%%%%%%%%%%%%%%%%%%%%%%
%%%%%%%%%%%%%%%%%%%%%%%%%%%%%%%%%%%%%%%%%%%%%%%%%%%%%%%%%%%%%%%%%%%%%%
The plate is seen to enhance the interatomic interaction in two
lobe-shaped regions to the left and right of atom $B$. This implies
that for a thin slab of an atomic gas at distance $z$ from the
plate, the atom--atom correlation function will be enhanced at
interatomic distances $r\simeq 2.5 z$ corresponding to the centres of
the lobes. By virtue of scale invariance, this holds for all $z$
that are compatible with the long-distance limit.
%
%%%%%%%%%%%%%%%%%%%%%%%%%%%%%%%%%%%%%%%%%%%%%%%%%%%%%%%%%%%%%%%%%%%%%%
%%%%%%%%%%%%%%%%%%%%%%%%%%%%%%%%%%%%%%%%%%%%%%%%%%%%%%%%%%%%%%%%%%%%%%
%
\paragraph{Summary and perspective.}
By considering the scaling behaviour for the respective Green
tensors, we have derived universal scaling laws for dispersion
interactions in the long- and short-distance limits as summarised in
Tab.~\ref{Tab1}.
%%%%%%%%%%%%%%%  T A B L E %%%%%%%%%%%%%%%%%%%%%%%%%%%%%%%%%%%%%%%%%%%
\begin{table}[!t!]
\begin{center}
\begin{tabular}{cccc}
\hline
 Distance
 &Long
 &\multicolumn{2}{c}{Short}\\ 
 Bodies&Magnetoelectric&Electric&Magnetic\\
\hline 
$U(\vect{r}_A)$&$1/a^4$&$1/a^3$&$1/a$\\
$U^{(0)}(\vect{r}_A,\vect{r}_B)$&$1/a^7$&$1/a^6$&$1/a^6$\\
$U^{(1)}(\vect{r}_A,\vect{r}_B)$&$1/a^7$&$1/a^6$&$1/a^4$\\
$\ten{T}(\vect{r})$&$1/a^4$&$1/a^3$&$1/a^3$\\ \hline
\end{tabular}
\end{center}
\caption{
\label{Tab1}
Scaling laws for the CP potential, free-space and body-induced vdW
potentials and the Casimir pressure.}
\end{table}
%%%%%%%%%%%%%%%%%%%%%%%%%%%%%%%%%%%%%%%%%%%%%%%%%%%%%%%%%%%%%%%%%%%%%%
Scaling laws indicate the absence of a characteristic length scale of
the system. For dispersion potentials, the typical interatomic
distances and the wavelengths of atomic and body response functions
give two such characteristic length scales. The nonretarded scaling
laws are hence only valid for distances well between these two length
scales while the long-range one is restricted to distances well above
the latter. 

The scaling laws may be used to deduce the functional dependence of
dispersion forces in the case where they depend on only a single
parameter. In more complex cases, the knowledge of a potential for a
body of given size can be used the infer the potential for a similar
body of different size. In particular, equipotential lines are
invariant under a scale transformation. More complex applications
include bodies with surface roughness. The duality invariance of
dispersion forces \cite{0827} can be used to extend our results to
magnetic atoms.

%%%%%%%%%%%%%%%%%%%%%%%%%%%%%%%%%%%%%%%%%%%%%%%%%%%%%%%%%%%%%%%%%%%%%%

\acknowledgments
This work was supported by the Alexander von Humboldt Foundation,
the UK Engineering and Physical Sciences Research Council, the SCALA
programme of the European commission and the CNRS.

%%%%%%%%%%%%%%%%%%%%%%%%%%%%%%%%%%%%%%%%%%%%%%%%%%%%%%%%%%%%%%%%%%%%%%

\end{document}